\documentclass{article}
\begin{document}
\title{Elimination of High-Energy Divergence in Relativistic Lagrangean Formulation of Gravitating Particle Dynamics}
\author{Anatoli Vankov}
\date{vankova@bethanylb.edu; anatolivankov@hotmail.com}

\maketitle

\section{Introduction} 
Different aspects of the gravitational problem are discussed, for example, in \cite{1} with further references. There is a statement that Newton's formulation of the gravitational law in 3-space (i=1, 2, 3) 
\begin{equation} 
\frac{d^2x^i}{dt^2}=-\frac{\partial\Phi}{\partial x^i}
\label{1}
\end{equation}
with the field equation
\begin{equation}
\nabla ^2\Phi=4\pi G\rho 
\label{2}
\end{equation}
cannot be incorporated into Special Relativity Theory (SRT). The Newtonian field propagates with infinite velocity, and one might expect that this assumption would be automatically corrected in the relativistic generalization of the gravitational law. However, numerous attempts to do this led to inherent contradictions or apparent inconsistence with experiments; for example, new relativistic gravitational field concepts resulted in the absence of bending of light in the gravitational field of Sun unless the General Relativity Theory (GRT) concept of curved space-time metric was introduced. At the quantum level, the gravitational problem remains unresolved; it discussed, for example, in \cite{2,3}. We think that such problems as the alleged incompatibility of gravity phenomenon with SRT, the nonremovable divergence in GRT field concept, the need of artificial  renormalization procedure in current field theories, and a general problem of gravitational field quantization point to the same root: the assumption of proper mass constancy in current field theories. In the present paper the problem of $1/r$ singularity of gravitational potential is investigated in the framework of Lagrangean formulation of Relativistic Dynamics of point particle with the proper mass dependent on field strength. We start with the introduction of the new mass concept. 

\section{The proper mass as a dynamical variable in the 4-coordinate and 4-momentum space}

Assuming the proper mass being field-dependent, let us consider the Larangian $L=m-W$ characterizing a point particle motion along the world line $s$. The Lagrangian consists of the two parts corresponding to proper kinetic and potential energy change; in other words, the change of the proper mass of a moving particle is due to the force field action. The Lagrangian may be considered in both 4-coordinate $x^\mu=(c_0t, x^i)$ ($\mu=0, 1, 2, 3$), and 4-momentum $p^\mu=mu^\mu$, where the proper velocity is $u^\mu{dx^\mu}/{ds}$. The zero component of any 4-vector is called a time part of the vector, while the space part is given by $\mu=i=1, 2, 3$. The scalar product is defined $x^\mu x_\mu=(x^0)^2-(x^i x_i)$ and similarly for any pair of 4-vectors. The world line element $ds$ relates to the proper time $\tau$ of the particle: $ds=c_0d\tau$ where $c_0$ is the speed of light at infinity. One should take into account that a massive particle travels along the time-like world line characterized by the independent equation: 
\begin{equation}
\frac{dx^\mu}{ds}\frac{dx_\mu}{ds}=-1
\label{3}
\end{equation}
Thus, in the above formulation of dynamical problem we deal with proper physical quantities in the 4-coordinate space, such as the proper mass, the proper time, the proper velocity, the proper momentum, the proper force, and so forth; they are measured in a comoving reference frame. Further, the particular case of the Lagrangian independent of the proper velocity is considered. Consequently, the equation of motion along the world line from the Hamilton Principle is
\begin{equation}
\frac{dm}{ds}-\frac{dW}{ds}=0
\label{4}
\end{equation}
There should be a relationship between the potential energy term and the tangent component of the force 4-vector $K^s=-{dW}/{ds}$. The proper velocity 4-vector $u^\mu$ is a unit vector tangent to the world line; therefore, $K^s=K^\mu u_\mu={\bf K\cdot \bf u}$, and the equation (\ref{4}) becomes 
\begin{equation}
\frac{dm}{ds}=-\bf{K}\cdot\bf{u}
\label{5}
\end{equation} 
Taking into account (\ref{3}) and following from it identity ${\bf u}\cdot({d{\bf u}}/{ds})=0$, it can be transformed into
\begin{equation}
{\bf u}\cdot{\bf u}\frac{dm}{ds}+m{\bf u}\cdot\frac {d\bf u}{ds}={\bf K}\cdot{\bf u}
\label{6}
\end{equation}
or with the 4-momentum ${\bf p}=m\bf u$
\begin{equation}
{\bf u}\cdot\frac{d{\bf p}}{ds}=\bf{u}\cdot\bf{K}
\label{7}
\end{equation}
Finally, the equations of motion in 4-coordinate form are
\begin{equation}
\frac{d}{ds}(mu^\mu)=K^\mu
\label{8}
\end{equation}

This is quite understandable because the tangent component of $\bf p$ is ${\bf p}\cdot {\bf u}=-m(s)$. Thus, we have a relativistic analog of the second Newton's law, in which the ordinary force is replaced by the Minkowski force $K^\mu$. The equations of motion (\ref{8}) are formulated in a covariant form in the 4-coordinate space. Similar ones are known in Relativistic Mechanics with the assumption of proper mass constancy. The assumption is discussed in \cite{4}); it is tacitly accepted in current relativistic field theories and is widely believed to be physically true, though the alternative was actually never investigated. The statement of this work is that Relativistic Gravitational Mechanics of point particle with the proper mass being a dynamical variable (the Alternative Relativistic Mechanics) is a natural relativistic generalization of corresponding Newton's gravitational dynamics. The usual way of deriving (\ref{8}) with the proper mass being field independent is to consider the relativistic Lagrangian in $(x^i, t)$ coordinate system for a single particle acted on by a conservative force independent of velocity \cite{5}; the goal is not to find equations of motion but rather chose a function $L$ for which the Euler-Lagrange equations, as obtained from the variational principal $\delta\int_{t_1}^{t_2}L\,dt=0$, ``agree'' with the known relativistic equations. The following Lagrangian seems to satisfy this requirement: $L=-mc_0^2\sqrt{1-\beta^2}-V$, where $\beta^2=v^iv_i/c_0^2$ is a squared magnitude of a coordinate 3-velocity $v^i={dx^i}/{dt}$, and $V$ is the potential depending only on position; a space part of the corresponding Minkowski force follows from the above form. Clearly, the factor $\sqrt{1-\beta^2}$ in $L$ appeared as a result of a coordinate transformation in action $\int{L}ds\to\int{L}c_0dt/\gamma$. The equations of motion $m\gamma {dv^i}/{dt}=F^i$ derived from the above Lagrangian are the ones following from the space part of the equations (\ref{8}) with the proper mass fixed. However, the role of time part of the Minkowski force here is not clear. Our view of the problem is, as follows. The variational principle, if applied to the problem of motion in 3-space and time, is not equivalent to that in the originally covariant form describing a motion along the world line in terms of ``running'' proper time $\tau$. These are different physical problem formulations having different solutions: the proper mass constancy in conventional Relativistic Mechanics should be considered a weak-field approximation of a general covariant solution, which includes five unknown functions $x^\mu(s)$, $m(s)$ obtained from five equations (\ref{8}) and (\ref{3}), as discussed in (\cite{4}). For example, in the conventional formulation the Minkowski force and momentum (or velocity) 4-vectors are always orthogonal. In the alternative solution, as is seen from (\ref{5}), the orthogonality may take place in a particular case when the term ${dm}/{ds}$ vanishes; the conditions can be found from the following equation equivalent to (\ref{6}):
\begin{equation}
{\bf u}\frac{dm}{ds}+m \frac{d{\bf u}}{ds}=\bf K 
\label{9}
\end{equation}
   
Let us consider now the Lagrangian $L=m-W$ in the 4-momentum space
$p^\mu=mu^\mu$, which is a complementary space, having space and time parts $p^\mu=(m, \gamma m v^i/c_o)$. Here, the proper mass $m$ plays the same role as the proper arc length $s=c_0\tau$ in the coordinate space; the proper 4-velocity should be obtained as $u\mu(p)={dp^\mu}/{dm}$, where the argument $\small p$ indicates that the velocity is defined in the momentum space; not surprisingly, the velocity $u\mu(p)$ coincides with the expression for 4-velocity in the 4-coordinate space $u^\mu(p)={dx^\mu}/{ds}$. Thus, we have $L=m-W(m)$ and want to find the trajectory of particle motion in the 4-momentum space; the solution should describe the potential energy as a function of the proper mass change under force action. Under earlier formulated conditions, we have a general solution ${dL}/{dm}=0$, or $dm=dW$. Taking the potential energy being zero and the proper mass being constant $m=m_0$ at infinity, we conclude that the proper mass should be connected to the potential energy in the following way:
\begin{equation}
m=m_0-W(m).
\label{10}
\end{equation}
The concrete dependence of mass on the strength field $m(x^i)$ should be found by solving equations (\ref{8}). It is important to note that the alternative Lagrangean problem formulation leads to a generalization of the conventional Relativistic Mechanics; therefore, the property of proper mass variability in the force field follows from the first principle independent of the type of force, and it is falsifiable. The variable proper mass plays a decisive role in the elimination of the so-called gravitational self-energy divergence problem; it will be illustrated by using the equation (\ref{10}) in the case of scalar potential $1/r$.      

\section{The gravitational divergence problem and new symmetries}
For the practical use of the above results, one should express the trajectory of particle motion (\ref{8}) in terms of 3-space coordinates depending on time $x^i(t)$ taking into account that $ds=c_0 dt/\gamma$, and the 4-vector velocity and momentum $x(t), t$ can be presented in the form $u^\mu=(\gamma,\gamma v^i/c_0)$ and  $p^\mu=(\gamma m,\gamma m v^i/c_0)$. The space part of the equations is  
\begin{equation}
\frac{d}{dt}(\gamma mv^i)=F^i 
\label{11}
\end{equation}
with the relationship between Minkowski and ordinary forces in 3-space
\begin{equation}
F^i={\frac{c_0^2}{\gamma}}K^i
\label{12}
\end{equation}
while the time part contains the time component of Minkowski force $K^0$:
\begin{equation}
\frac{d}{dt}(\gamma m)=\frac{c_0}{\gamma}K^0
\label{13}
\end{equation}
The equation (\ref{13}) reflects the energy balance and can be expressed in terms of space part of Minkowski force from (\ref{5}):  
\begin{equation}
\frac{d}{dt}(\gamma m c_0^2)=F^iu_i+\frac{c_0^2}{\gamma}\frac{dm}{dt}
\label{14}
\end{equation}
In (\ref{14}) the term $\frac{c_0^2}{\gamma}{dm}/{dt}$ describing the role of the proper mass variation in relativistic dynamical process is recovered (probably, for the first time in practice of Relativistic Mechanics applications). 

To investigate a free fall problem, we have to solve the equations (\ref{11}) and (\ref{14}) in the case of static spherical symmetric gravitational field due to a point source of mass $M$. The gravitational force acting on a test particle of variable proper mass $m(r)$ at a distance $r$ from the center is $F_g(r)dr=c_0^2m(r)d(r_g/r)$, where $r_g=GM/c_0^2$ is ``the gravitational radius''. For simplicity, the case of the test particle having zero total energy at infinity is considered. The following denotations are used: $dr=vdt,\ 1/\gamma=\sqrt{1-\beta^2},\ \beta=v/c_0, \ 1/\gamma_r=1-r_g/r$. The total energy conservation requires that the time component of Minkowski force $K^0$ and the corresponding left-hand part of (\ref{14}) is zero, $E=\gamma m c_0^2=m_0c_0^2$, and the energy balance is described by the equation
\begin{equation}
p(r)^2+m(r)^2c_0^2=m_0^2c_0^2
\label{15}
\end{equation}
where the momentum is $p(r)=m_0v(r)$ and $m(r)/m_0=1-r_g/r$; it is convenient to present the equation in the form
\begin{equation}
m_0^2\beta^2+ m(r)^2=m_0^2
\label{16}
\end{equation}
It is seen that $\gamma_r=\gamma$ that means that the gain of kinetic energy is due to the change of potential energy $W=m(r)c_0^2(r_g/r)$. Because the proper mass cannot be negative, the above equations have physical sense within the range $r\ge r_g$. At $r=r_g$ the proper mass and the baryon charge should vanish; this does not look physically possible, that is why we consider the results in the range $r>r_g$. It is interesting to note that in the case of {\it static potential} when the particle is moved by some transporting device with a small constant speed, the proper mass changes exponentially, as is seen from (\ref{14}):
\begin{equation}
m(r)/m_0=exp(-r_g/r)
\label{17}
\end{equation}
and a particle theoretically can approach the center within a limit $r/r_g\to 0$. The dependence of a relative velocity in a central fall from rest at infinity is found from (\ref{16}):
\begin{equation}
\beta=[1-(m/m_0)^2]^1/2=[1-(1-r_g/r)^2]^1/2        
\label{18}
\end{equation}
As was emphasized, it is valid in the range $r>r_g$.

The result of fundamental importance is that the problem of $1/r$ gravitational potential divergence is eliminated in the SRT Mechanics framework due to the effect of ``proper mass exhaustion''. The effect can be treated as a test particle back-reaction on the source action; in the conservative force field the total energy of the particle is limited to the maximal value stored in the proper mass (plus kinetic energy, if any) at infinity. Consequently, potential energy is proportional to the variable proper mass convertible to kinetic energy. In the case of particle-particle interaction, the total energy of the system include the energy stored in variable proper masses of both particles. Thus, this is the assumed constant proper mass that causes the ``self-energy'' divergence rather than the $1/r$ potential. The variable proper mass concept is justified by considering the Lagrangean formulation of Relativistic Mechanics in both coordinate and momentum complementary spaces similarly to Quantum Mechanics methodology. The scalar product of vectors $d{\bf x}=(d{\bf x}/ds)ds$ and ${\bf p}$ is 
\begin{equation}
d{\bf x}\cdot{\bf p}= -c_0 m(s)d\tau(s)
\label{19}
\end{equation}
Recalling a relationship between the proper energy to the quantum frequency (or to the proper time interval) $m(s)c_0^2=h/d{\tau}$, one can conclude that the known observed effect of gravitational time dilation is caused by the roper mass dependence on field strength. The product is invariant in the gravitational field $d{\bf x}=(d{\bf x}/ds)ds=-h/c_0$; it shows the intimate relationship between the Plank's constant in a quantum theory and the ultimate speed of light in a relativistic theory, and a necessity of a relativistic quantum theory of gravitational field.

Our gaining into insight of the problem formulation in both coordinate and complementary (momentum) space reveals field conservation properties understandable only at the new relativistic level. As is seen from (\ref{15}), the total energy conservation law means the constancy of the improper mass $\gamma(r)m(r)=m_0$ of a particle in free fall (from rest at infinity) and reflects the symmetry of a momentum real rotation in the $(p^i$, $m$) momentum space. The same rotational symmetry takes place in the $(x^i$, $t$) coordinate space due to constancy of the improper time interval $c_0dt(r)=\gamma(r)d\tau(r)=\tau_0$. This is seen from the metric form of the world line interval of the particle motion in the above case $ds^2=c_0^2d\tau^2=c_0^2dt^2-dr^2$. Instead of conserved improper total mass $\gamma(r)m(r)=m_0$ we have the corresponding conserved improper time interval $c_0dt(r)=\gamma(r)d\tau(r)=\tau_0$ (as measured at infinity in a source-centered frame). The conservation is justified by the quantum connection $\gamma m c_0^2=h\gamma/dt=h/ d\tau $, from which it follows $m(r)\to m_0$ and $\tau(r)\to \tau_0$ at $r\to \infty$. In addition to (\ref{15}), we have  
\begin{equation}
dr^2+d\tau^2c_0^2=d\tau_0^2c_0^2
\label{20}
\end{equation}
Thus, measurements of the improper mass and the improper time of a particle in free fall does not reveal the presence of field unless the corresponding proper quantities are measured and compared.

\section{Discussions and conclusion}
We challenge a general opinion, discussed, for example, in \cite{1}, that the gravity phenomenon is incompatible with the SRT framework. In our alternative SRT-based approach, the current mass-energy concept and, correspondingly, the field concept is subject to revision. Looking at the equations (\ref{1}), one can realize that a relativistic generalization of Newton's formulation of the gravitational law is indeed impossible within the concept of constant proper mass. A probing a field means the introduction of a test particle; the mass of the test particle is not present in (\ref{1}). Our finding is that the $1/r$ gravitational potential singularity in a field source is not removable by itself: this is the potential energy that is physically free of divergence provided the proper mass is a dynamical variable consistently defined in the Lagrangean (alternative) formulation of Relativistic Mechanics. The equation of field due to a system of interacting particle (the source) in (\ref{2}) becomes nonlinear because the field causes a proper mass defect characterizing a binding energy of the system. Similarly, the test particle is affected by the field: its proper mass reduction characterizes field strength. In other words, the standard proper mass at a given point is the field. In this theory, the solution of dynamical equation of test particle motion (as shown in the example of free central fall) is free of divergence. 

The question arises: why a renormalization procedure of divergence elimination occured to be successful in electromagnetic theory but does not work in a gravitational theory? To answer this question, one should compare the gravitational potential energy $W_g=m_0C_0^2(r_g/r)$ (with the gravitational radius $r_g=GM/m_0C_0^2$) and the equivalent electric (Coulomb) energy $W_e=m_0c_0^2(r_a/r)$ (with the annihilation radius $r_a=kQq/m_0C_0^2$): both ultimately are $m_0C_0^2$. It is seen that the Coulomb potential can be ``exhausted'' at a distance $r=r_a>>r_g$ characterizing a particle ``electric'' size while the release of gravitational energy is physically impossible in particle interactions: relativistic gravitational physics is subjected to astrophysical conditions of big mass objects. This explains why some artificial procedure of ``cutting off'' high momentum harmonics arising from distances less than $r_a$ might solve (but not resolve) the problem of electromagnetic field divergence. We think that the need for ``renormalization'' in a field theory is a manifestation of problem to be resolved at the fundamental level.

The scope of this work does not allow us to discus perspectives of the alternative approach for the development of relativistic electromagnetic divergence-free theory, neither we are able to discuss here in detail issues related to the consistency of the alternative approach with experiments: these topics need special works. For further our materials, a reader is referred to electronic preprints: gr-qc/0311063, ``On the problem of mass origin and self-energy divergence in relativistic mechanics and gravitational physics''; gr-qc/0105057, ``Proposal of experimental test of general relativity theory''; PhilSci Archive ``On the possibility of motion with the speed greater then the speed of light'' (February, 2003). The main purpose of the present work is to discuss the idea of the alternative approach to the relativistic field theory development. The central claim is that the variable proper mass concept is a natural generalization of the current concept of a constant proper mass (as assumed in GRT, Relativistic Electrodynamics and quantum field theories); the alternative concept seems to be perspective for the principle elimination of the field divergence problem. Let us briefly sum up some preliminary results concerning experimental tests.

Obviously, the alternative approach is falsifiable and needs to be verified by relativistic gravitational experiments, such as measurements of Mercury orbit precession, gravitational red shift, time dilation, and radar echo time delay. It should be emphasized that reliable experimental data relevant to the problem were obtained under weak-field conditions. Our rigor metric analysis showed that the Schwarzschild metric form in GRT and the one in the alternative approach coincide {\it under weak-field and low-energy conditions}; the difference is in physical interpretations of calculated results. Consequently, the predictions are numerically the same in a leading (linear) term, what means that one cannot distinguish between the two theories by the criterion of the above mentioned and similar experiments. Gravitational properties of photon were tested in some of them. In our mass-energy concept, the total energy conservation requires the constancy of photon frequency along with a momentum change for the photon in gravitational field: the field acts on the photon as a refracting medium does. Thus, the proper mass dependence on field strength together with our photon concept gives a physical explanation of the experiments involving the photon in terms of gravitational potential, as opposed to the GRT treatment in terms of space-time curvature. There was a suggestion (\cite{6}) to introduce a dependence of atomic frequency on static gravitational potential together with a conservation of the photon total energy in the GRT interpretation of the gravitational red shift; however, the effect is consistently treated in GRT in terms of space-time curvature, and the suggestion is in contradiction with the GRT foundations: the proper mass constancy is clearly embedded in the GRT field equations, while such notions as a force, kinetic and potential energy are not relevant there. As was mentioned before, there were unsuccessful attempts to develop a gravitational theory within the SRT framework in some post-Newtonian model of photon being ``attracted'' by a gravitational force; the model was supposed to explain the deflection of light grazing the Sun. One of the arguments for an exclusion of the gravity phenomenon from the SRT domain came after realizing that a requirement of coupling the photon to a gravitational scalar field contradicts to the fact that the stress-energy tensor for the electromagnetic field in Minkowski space is traceless (that is, not allowing the coupling). Our finding that the photon deflection is due to the gravitational refraction rather than the force attraction resolves this issue. It is interesting to note that the gravitational refraction effect was discussed in (\cite{7}) from a different viewpoint.    

A difference in our and GRT predictions of strong-field effects rises with field strength. As an instructive example, let us compare our and GRT predictions of radial motion of a photon and a massive particle in a spherical symmetric gravitational field  outside a massive sphere (Earth, for example), field strength being characterized by the gravitational radius $r_g$. In both theories, an observer ``at infinity'' in a sphere-centered reference frame should detect the photon slowing down when approaching Earth. As for the particle, predictions are different. According to GRT, the particle also should slow down so that its speed could not exceed the photon speed no matter how great the initial kinetic energy (or initial Lorentz factor $\gamma_0$ at infinity was. The point is that for $\gamma_0$ exceeding the ``critical'' value $\gamma_0\ge {3/2}$, the particle would only decelerate (as if some ``resisting'' force overcoming the gravitational one was exerted on it). In our approach, the particle in this example would only accelerate (regardless of initial speed) consistently with the conservation law of potential-to-kinetic energy transformation; the gravitational properties of the photon were above explained within the same concept. From this, it follows that the particle having initial energy above some threshold $\gamma _{thr}$ should become {\it superluminal} what is impossible in GRT. This prediction of a new physical phenomenon does not depend on a metric consideration: the question whether the gravitational superluminocity exists can be experimentally verified. Our calculations show that ultra-high energy particles falling onto Earth with $\gamma _{thr}>2\cdot 10^4$ are superluminal. Such particles are present in cosmic rays; therefore, their superluminocity could be tracked by Cherenkov radiation. The latter has to be very specific: photons are emitted backward in a narrow cone (photon flashes seem coming from Earth), while flash durations should be statistically distributed in $ms$ range (typical time of photon flight upward to a detector aboard an experimental satellite); count statistics should fit cosmic rays flux above the threshold for given geometrical and physical detection efficiency. Our proposal of the new gravitational test is the topic of another work. 

The main conclusion of this work can be formulated in the statement that the so-called self-energy divergence due to $1/r$ gravitational potential is shown to be eliminated in the concept of field-dependent proper mass. Further investigations of the problem are needed to understand a perspective of the concept for the development of divergence-free relativistic field theories.

\end{document}